\documentclass[twocolumn,showpacs,preprintnumbers]{revtex4}
\usepackage{amsmath}
\usepackage{epsfig}
\usepackage{graphicx}
\usepackage{rotating}
\usepackage{amssymb}
\usepackage{mathptmx}
\usepackage{amsmath}
\usepackage{float}
\usepackage{natbib}

\begin{document}

\title{Finite size effect on the specific heat of a Bose gas in multifilament cables}
\author{G. Guijarro}
\affiliation{Posgrado en Ciencias F\'isicas, UNAM; Instituto de F\'isica, UNAM}
\author{M. A. Sol\'{\i}s}
\affiliation{Instituto de F\'{\i}sica, UNAM, Apdo. Postal 20-364, 01000 M\'exico, D.F., MEXICO }

\begin{abstract}
The specific heat for an ideal Bose gas confined in semi-infinite 
multifilament cables is analyzed. We start with a Bose gas inside a semi-infinite tube of impenetrable walls and  finite rectangular cross section.  The internal filament structure is created by applying to the gas two, mutually perpendicular, Kronig-Penney delta-potentials along the tube cross section, while particles are free to move perpendicular to the cross section. The energy spectrum accessible to the particles is obtained and introduced into the grand potential to calculate the specific heat of the system as a function of temperature for different values of the  periodic structure parameters such as: the cross section area, the wall impenetrability and the number of filaments. The specific heat  as a function of temperature shows at least two maxima and one minimum.
The main difference with respect to the infinite case is that the peak associated with the BE condensation becomes a smoothed maximum or in other words, there is not a jump in the specific heat derivative, whose temperature no longer represents a critical point.
\end{abstract}
\maketitle
\section{Introduction}
Since the discovery of the superfluidity in helium four in 3D \cite{superfluidez}, a lot of question appeared around the existence of superfluidity in low dimensional systems. Thenceforth and to date many experimental \cite{Gasparini,nanotubes} and theoretical \cite{canadiense,Pathria77} studies have been realized to understand the dimensional and finite size effect on superfluidity and Bose-Einstein condensation (BEC).

Although BEC in three-dimensional infinite systems takes place at a well-defined critical temperature $T_0$, for finite or semi-infinite systems the corresponding ``transition" extends over a finite range of temperatures, around $T_0$ which becomes only a reference temperature. An extensive study was carried out by Pathria and collaborators \cite{Pathria77} who eventually proposed that the temperature at the maximum of the specific heat should be the critical temperature for a Bose gas within a semi-infinite slab.

In this article we study the specific heat of an ideal Bose gas within a semi-infinite periodic array of filaments. We want to give a detailed description only of the effects of the periodic structure over the bosons, so we are not taking into account any kind of interactions between particles. We calculate the specific heat as a function of the cross section size, the number, and the wall impenetrability (or penetrability), of the filaments placed neatly as a bundle to form a cable.  
The occurrence of more than one maximum in the specific heat, leads us to reconsider Pathria's ``critical temperature" for finite and semi-infinite system. 

\section{Multifilament structure}
We consider N non-interacting bosons confined in a periodic array of filaments with finite rectangular cross section and infinite length, set together to form a cable whose external wall is impenetrable. The filament structure is created by applying to the gas two set of Dirac deltas mutually perpendiculars along the $x$ and $y$ directions while the particles are free to move along the $z$ direction, i.e., finite numbers $M_x$ and $M_y$ of identical deltas, which are separated from each other a fixed distance $a_i$ $(i=x \ \mbox{o} \ y)$, are applied in the $x$ and $y$ directions, respectively. Thus, particles are confined in a box of impenetrable potential with dimensions: $L_x=a_x(M_x+1)$, $L_y=a_y(M_y+1)$ and $L_z=\infty$. The potential expression is  
\begin{equation}
\label{eq01}
V(x,y)=\sum_{m_x=1}^{M_x} v_x \delta(x-m_xa_x) + \sum_{m_y=1}^{M_y} v_y \delta(y-m_ya_y) 
\end{equation}
where $v_x$ y $v_y$ are the intensities of the delta potential in the $x$ and $y$ directions, respectively. The Schr\"odinger equation for each boson of mass $m$ in the system is
\begin{equation}
\label{eq02}
\left[ -\frac{\hbar^2}{2m}\nabla^2+V(x,y) \right] \Psi(x,y,z)=\varepsilon_k \Psi(x,y,z)
\end{equation}
which is separable in each direction $\Psi(x,y,z)={X}(x){Y}(y){Z}(z)$, such that $\varepsilon_\bold{k}=\varepsilon_{k_x}+\varepsilon_{k_y}+\varepsilon_{k_z}$ is the energy per particle. In the $z$ direction the energy is given by 

\begin{equation}
\label{eq03}
\varepsilon_{k_z} =\frac{\hbar^2 k_z^2}{2m}
\end{equation}
with $k_z=2\pi n_z/L_z$ the wavenumber in the $z$ direction, $n_z=0,\pm1,\pm2,...$ due to the periodic boundary conditions in a box of length $L_z$. The boundary conditions used in the $x$ and $y$ directions are ${X}(0)={X}(L_x)=0$ and ${Y}(0)={Y}(L_y)=0$, that is, the wave function vanishes at the walls of the tube. The energies of the particles in $x$ and $y$ directions are implicitly obtained from the equations \cite{KP,Griffiths}
\begin{equation}
\label{eq04}
P_{0i} a_{0i}\frac{\sin(\alpha_i a_i)}{\alpha_i a_i}+ \cos(\alpha_i a_i)=\cos\left(\frac{n_i\pi}{M_i+1}\right) 
\nonumber
\end{equation}
with $n_i = 1, 2, ..., M_i+1$, where $\alpha_i^2\equiv2m\varepsilon_{k_i}/\hbar^2$ $(i=x \ \mbox{o} \ y)$, $M_i$ is the number of delta potentials. We rewrite the dimensionless constants $Pi=mv_ia_i/\hbar^2$ as $P_i=(mv_ia_i\lambda_0/\hbar^2)(a_i\lambda_0)\equiv P_{0i}(a_i/\lambda_0)$, where $\lambda_0 \equiv h/\sqrt{2 \pi mk_BT_0 }$ is the de Broglie thermal wavelength of an ideal Bose gas in an infinite box at the critical temperature $T_0=2 \pi\hbar^2\rho^{2/3}/mk_B \zeta(3/2)^{2/3} \simeq  3.31\hbar^2\rho^{2/3}/mk_B$, with $\rho\equiv N/L^3$ the boson number density and $a_i$ the distance between the delta barriers along the $i=x$ and $y$ directions. $P_{0i} \equiv mv_i\lambda_0/\hbar^2$ is a measure of the tube wall impenetrability directly related to the delta-barrier strength $v_i$ and $a_{0i} \equiv a_i/\lambda_0$.

\section{Specific heat}
The thermodynamic properties are obtained from the grand potential $\Omega(T,V,\mu)$, for a Bose gas is given by the expression 
\begin{equation}
\label{eq05}
\Omega(T,V,\mu)=k_BT \sum_{\bold{k}} \ln[1-e^{-\beta(\varepsilon_{\bold{k}} - \mu)}]
\end{equation}
The thermodynamic properties can be calculated using the relations
\begin{equation}
\label{eq06}
\begin{aligned}
N=-\left(\frac{\partial \Omega}{\partial \mu}\right)&_{T,V} ,  \quad    U=-k_BT^2\left[\frac{\partial}{\partial T}\left(\frac{\Omega}{k_BT}\right)\right]_{V,z}\\
& \mbox{and}\quad C_{V}=\left(\frac{\partial U}{\partial T}\right)_{N,V}\\
\end{aligned} 
\end{equation}
where $N$ is the particle number, $U$ is the internal energy, $C_V$ is the isochoric specific heat and $z \equiv e^{\beta \mu}$ is the fugacity. To find the chemical potential $\mu$ we begin with the number equation
\begin{equation}
\label{eq07}
N=\sum_{k_x,k_y}L_z\left(\frac{m}{2 \pi \hbar^2 \beta}\right)^{1/2}g_{1/2}(z_0)
\end{equation}
where the integral over infinite $z$ direction has been done and we have introduced the Bose functions\cite{PathriaBook} $g_\sigma (t)\equiv \sum_{l=1}^{\infty}t^l/l^\sigma$.
The boson number density of the system is
\begin{equation}
\label{eq08}
\frac{N}{L_xL_yL_z}=\sum_{k_x,k_y}\frac{1}{a_{x}(M_x+1)a_{y}(M_y+1)}\left(\frac{m}{2\pi \hbar^2 \beta}\right)^{1/2}g_{1/2}(z_0)
\end{equation}
In order to use the critical temperature $T_0$ as a reference unit, here we set the boson number density $N/L_xL_yL_z$ equal to that of an ideal Bose gas in the thermodynamic limit. After that, we find the chemical potential as a function of temperature for the calculation of other thermodynamic properties such that internal energy and specific heat.

From (\ref{eq05}) and (\ref{eq06}) we obtain an expression for the isochoric specific heat 
\vspace{-0.1cm}
\begin{equation}
\begin{aligned}
\label{eq09}
&\frac{C_{V}}{Nk_B}=\frac{1}{a_{0x}a_{0y}\zeta(3/2)(M_x+1)(M_y+1)}\sum_{k_x,k_y}\left[ \frac{3}{4}\left(\frac{T}{T_0}\right)^{1/2}\times \right.\\
&g_{3/2}(z_0)+ \frac{1}{2}\left(\frac{T_0}{T}\right)^{1/2}\left(2\tilde{\varepsilon}_{k_x}+2\tilde{\varepsilon}_{k_y}+T\frac{\partial \tilde \mu}{\partial T}-\tilde \mu\right)g_{1/2}(z_0)\\
&\left.+\left(\frac{T_0}{T}\right)^{3/2}(\tilde{\varepsilon}_{k_{x}}+\tilde{\varepsilon}_{k_{y}})\left(\tilde{\varepsilon}_{k_x}+\tilde{\varepsilon}_{k_y}+T\frac{\partial \tilde\mu}{\partial T}-\tilde \mu\right)g_{-1/2}(z_0)\right]   \\ 
\end{aligned}
\end{equation}
with $\tilde {\varepsilon}_{k_i}\equiv\varepsilon_{k_i}/k_BT_0$ and $\tilde{\mu}\equiv\mu/k_BT_0$.
\vspace{-0.5cm}
\begin{figure}[H]
\begin{center}
\includegraphics[width=8.5cm,angle=0]{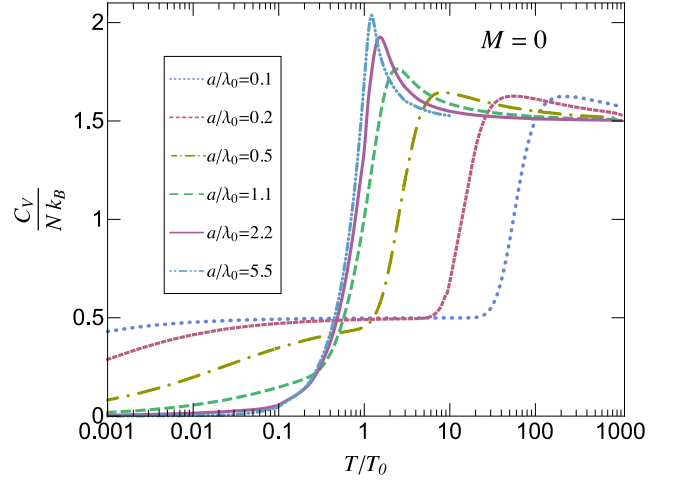}
\caption{Specific heat as a function of $T/T_0$, of a semi-infinite tube with different square cross section $(a/\lambda_0)^2$ values (Color figure online).}
\label{CvTubos}
\end{center}
\end{figure}
For the isotropic case, where $P_{0x}=P_{0y}\equiv P_0$, $a_x/\lambda_0=a_y/\lambda_0\equiv a/\lambda_0$ and $M_x=M_y\equiv M$, we plot in Fig. \ref{CvTubos} the isochoric specific heat as a function of temperature, of a semi-infinite tube of impenetrable walls, for different square cross section $(a/\lambda_0)^2$ values. As the temperature goes to zero $C_V\to0$. The specific heat present one maximum, then goes slowly as the temperature increases to the value $1.5$ as expected. For $a/\lambda_0=0.1$ and $0.2$ we observe a one-dimensional behavior over a relative large region of temperatures, that is, $C_V/Nk_B$ approaches the classical value $1/2$.  
\begin{figure}[H]
\begin{center}
\includegraphics[width=8.5cm,angle=0]{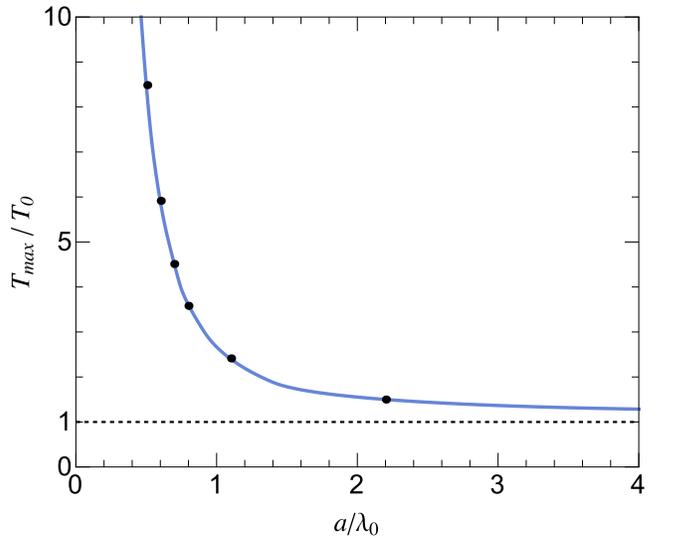}
\caption{Temperature $T_{max}$ at which the specific heat of a semi-infinite tube of square cross section $(a/\lambda_0)^2$ is maximum (Color figure online).}
\label{MaxM0}
\end{center}
\end{figure}

Let $T_{max}$ the temperature at which the specific heat is maximum. 
In Fig. \ref{MaxM0} we plot $T_{max}/T_0$ as a function of $a/\lambda_0$. As the cross section area of the semi-infinite tube increases $T_{max}$ decreases monotonically approaching to $T_0$ as expected. This result agrees with the ones obtained by R. K. Pathria \cite{Pathria72}, they analyzed an ideal Bose gas confined to a thin-film geometry $(\infty\times\infty\times D)$ under a variety of boundary conditions.

\begin{figure}[tbh]
\begin{center}
\includegraphics[width=8.5cm,angle=0]{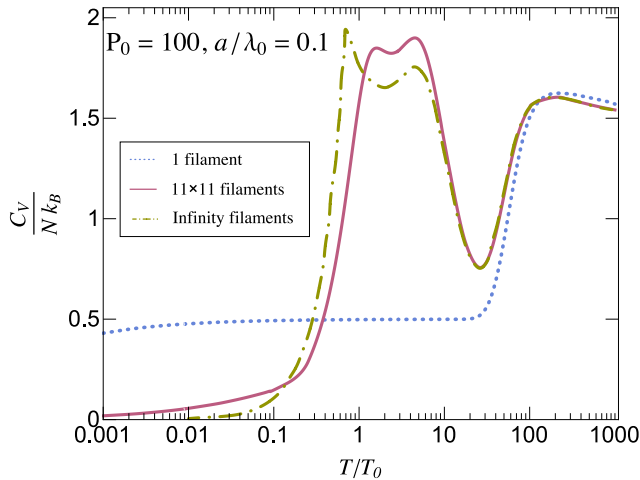}
\caption{Specific heat as a function of $T/T_0$ for different numbers of filaments of cross section $(a/\lambda_0)^2=(0.1)^2$ and $P_0=100$ (Color figure online).}
\label{figTubosa}
\end{center}
\end{figure}


The specific heat, for three-dimensional infinite systems, is know to possess a cusplike singularity at the critical temperature $T_0$. In order to explore finite size effects on periodic array of filaments three different systems are analyzed. First, we consider an impenetrable tube with finite cross section area $(a/\lambda_0)^2$ and infinite length (we call 1 filament). Second, a square array of 121 penetrables filaments of infinite length inside a hard tube is studied, this array in created by set $M=10$ ($11\times11$ filaments). The third system analyzed is a infinite periodic array of filaments of infinite length\cite{Paty} (infinite number of filaments). We remark that the filaments composing the three different systems have the same finite cross section area $(a/\lambda_0)^2$ and infinite length. 

In Fig. \ref{figTubosa}, we plot the isochoric specific heat as a function of $T/T_0$ for the three periodic array of filaments described above, with $a/\lambda_0=0.1$ and $P_0=100$. The effect of the internal filament structure (KP delta potentials) is evidenced by some maxima and minima present in the specific heat. The minimum, at larger temperature, is associated to particle trapping between two planes and appears at temperatures such that the thermal wavelength satisfies $\lambda\simeq 2\;a$. The first maximum, from lower to higher temperature, reflects the preamble of the appearance of a Bose-Einstein condensation. The maximum at the largest temperature marks la return of the system to its 3D free ideal Bose gas behavior. This maximum regularly appears when the wavelength of the Broglie is $\lambda\simeq 0.7\;a$. We do not have a detailed explanation of the origin of the maximum and minimum intermediates in  $C_V/Nk_B$ for $a/\lambda_0=0.1$.  

The maxima present in the system with infinite number of filaments also appear in the semi-infinite system, but the maxima are moved towards higher temperatures due a finite size affect. On the other side, the minimum associated with particle trapping coincide in both systems. The existence of a BEC is clear in the infinite system (infinite number of filaments) which is manifested by a peak at $T=T_c$ (critical temperature of the system with infinite number of filaments), but for the system with 121 filaments we have a smoothed maximum or in other words, there is not a jump in the specific heat derivative, whose temperature no longer represents a critical point.

%
\begin{figure}[tbh]
\begin{center}
\includegraphics[width=8.5cm,angle=0]{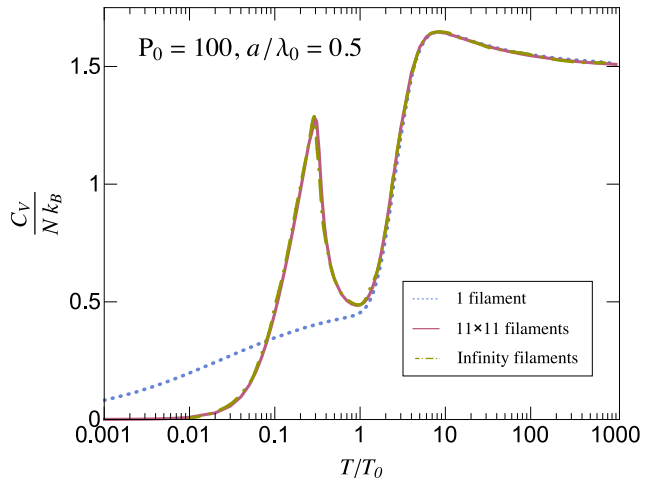}
\caption{Specific heat as a function of $T/T_0$ for different numbers of filaments of cross section $(a/\lambda_0)^2=(0.5)^2$ and $P_0=100$ (Color figure online).}
\label{figTubosb}
\end{center}
\end{figure}

Fig. \ref{figTubosb} shows the specific heat (in units of $k_BT$) as a function of $T/T_0$, for the three same arrays of filaments described above, but now we have increased the cross section area of each filament to the value $(a/\lambda_0)^2=(0.5)^2$. In this case, the extreme values preset in the specific heat for infinity filaments and $11\times11$ filaments mach and the only difference is in the critical temperature. The maximum associated with the return to the 3D behavior coincide for three system.

\section{Conclusions}

In conclusion, a detailed analysis of the isochoric specific heat of non-interacting bosons confined in semi-infinite multifilament cables was performed. We showed that the specific heat as a function of temperature has at least two maxima and one minimum. The first maximum, from lower to higher temperature, reflects the preamble of the appearance of a Bose-Einstein condensation. The temperature of this maximum is smaller as the cross section area or the wall impenetrability increases.
The maximum at the largest temperature marks the return of the system to its 3D free ideal Bose gas behavior. This maximum regularly appears when the wavelength of de Broglie is $ \lambda \simeq 0.7 a$. The minimum and/or plateau reflects the tendency of the system to behave in 1D. The minimum appears at temperatures such that $ \lambda \simeq 2 a$ becoming a plateau at lower temperatures 
for impenetrability such that $P_0 > 100$. The temperature $T_{max}$, at which the specific heat of a semi-infinite tube of square cross section $(a/\lambda_0)^2$ is maximum, decreases monotonically as $a/\lambda_0$ increases. As the cross section area approaches infinity $T_{max} \to T_0$ as expected. The maxima present in the specific heat for a infinite system also appear in the semi-infinite system, but the maxima are moved towards higher temperatures due a finite size affect. However when we have an array greater than 11 $\times$ 11 filaments with $a/\lambda_0=0.5$ the difference is only at the critical temperature. The main difference with respect to the infinite case \cite{Paty} is that the peak associated with the BE condensation becomes a smoothed maximum or in other words, there is not a jump in the specific heat derivative, whose temperature no longer represents a critical point.

%
%

\begin{acknowledgements}
We thank support from grants PAPIIT-IN111613 and CONACyT 221030.
\end{acknowledgements}


\end{document}